# Customised high-value document generation

Niek du Preez[1], Nicolas Perry[2], Alexandre Candlot[2], Alain Bernard (1)[2], Wilhelm Uys[1], Louis Louw[1]

[1]University of Stellenbosch – Industrial Engineering Department – South Africa

[2]IRCCyN (UMR CNRS 6597), Virtual Engineering for Industrial Engineering - Nantes - France

**Abstract**

Contributions of different experts to innovation projects improve enterprise value, captured in documents. A subset of them is the centre of expert constraint convergence. Their production needs to be tailored case by case. Documents are very often considered as knowledge transcription. As the base of a structured knowledge-based information environment, this paper presents a global approach that helps knowledge-integration tool deployment. Two contrasting examples indicate how fundamental understanding of domain infrastructure contributes to a more coherent architecture of knowledge-based information environments. The first example is related to insurance services and uses roadmap and ontology to redefine a more coherent insurance contracting relation. The second example focuses on aircraft manufacturing and develops through diagram modelling, application specifications for process plan definition.

## 1 INTRODUCTION

### 1.1 Documentation Complexity

Lean extended enterprise and build-to-order induce better integration of PLM activities that go through computer aided systems and knowledge-based information environments [1]. This change of landmarks from physical document to electronics claims to redefine information support functionalities [2].

Specifications or constraints are usually transmitted from one expert to the other in a global convergence [3]. The differences between their competencies limit the global understanding of problems. Computer integration in the expertise chain aims to optimise this kind of relations and thus the use of enterprise knowledge.

The number of different enterprise concepts and complexities caused by different interpretations of these concepts encourages enterprises to standardise concepts and formalise behaviours. These efforts build re-usable and adaptable platforms and imply deep business architecture redeployments [4]. The rapidly changing environment requires convenient collaboration knowledge integration tools [5] and interoperability between different information sources [6].

The main difficulty encountered is to control the complexity of information quantity and informality. A reduction of the work-structure diversity helps in this regard to optimise the information efficiency. Seen from the process of collaborative document writing, it is hard to guarantee the consistency of the document's subject. The contents of shared documents may deal with many subjects and fields, and every collaborator would compose the text by its own understanding, different from other's because of different knowledge backgrounds. As a result the subject of the shared documents will be inconsistent among many copies. Furthermore, the communications between collaborative systems, the systems and environments will also influence the consistency of the subject. An approach focussing on the semantic and syntax distinction can help to resolve the consistency problems involved in collaborative writing [7].

### 1.2 Project Context

This work is part of a collaboration project between two research teams supported by NRF in South Africa and CNRS in France. It aims at the identification of possible synergy around performance indicators for knowledge management improvement

This collaboration starts from a global observation. When benefits from productivity optimisation become harder to obtain, the market expectancies are changing faster. Enterprises have to analyse and control their core competencies to react efficiently to this new challenge. In the following sections, two different approaches on two different application fields prove the interest of a global methodology for the creation of information consolidation tools in order to build structured knowledge-based information environments.

This paper presents a three-phased methodology to optimise and ensure coherent enterprise documentation:

1. At first must be identified the fundamental elements of the structure. It corresponds to the Infrastructure Definition Phase.
2. Relationships between these elements are then identified, and the elements are deployed in a coherent manner to optimise their efficiency. It is the Architecture Phase.
3. The third phase is document generation by a validated knowledge-based application

## 2 KNOWLEDGE BASED COMPLEX DOCUMENT GENERATION

### 2.1 Understand Infrastructure

The enterprise Infrastructure is constituted of elementary concepts that can be classed among process, products, resources and external effects [8]. These concepts enable the specification of all enterprise objects relating to three different points of view: functional, behavioural and structural (FBS) [9].

Their boundaries can be retrieved through the perception ability of stakeholders [10]. Each person naturally does this division, but the formalisation of a common understanding is harder to accomplish. The reason is that knowledge and meaning can't be externalised from humans to computers [11] or other documents. Meaning contained in representations has to be internally rebuilt by users [10].

Documents are considered as the inscription of knowledge and the problem is to analyse and propose structuration concepts as a base for their management. Ontology is one of the possible ways to achieve this goal. Research on ontology [12] seeks to provide enterprises with concept definition and management tools [13] [14] [15]. The common main steps are:

- Domain limit definition
- Manual or Automatic Corpus Analysis
- Concept extraction and organisation

The aim of this first phase is to differentiate concepts. The analysis of their relationship is part of a second phase focusing on Architecture. Concepts "behave" differently according to their context. The modelling of this structure and the analysis of its possible evolution constitute the Architecture phase.

### 2.2 Construct the Architecture

The maturity of the Infrastructure knowledge leads to a restricted number of concepts. They are more relevant and meaningful for defining the specifics of the studied domain. They are usually formed by a general name (corresponding for example to UML Class) and a limited numbered of typological instances (corresponding to UML Object).

The sum of their behaviour constitutes an as-is platform from which all the business outputs are derived. Usually the platform is build informally according to the enterprise evolution. It raises incoherencies in concept levels or typology definitions. Concretely it can be illustrated by a misuse of a machine (unclear relationship between a process and a resource), an inefficient procedure (confusion in processes), or an unsatisfactory product (unconsidered external effects, badly defined core product concepts).

In order to optimise platform efficiency, the roles of the Architecture phase can be defined as follow:

- Ensure the coherency between concepts
- Optimise relationships and build working environments
- Evaluate model maturity and complexity reduction
- Define a coherent integration method of knowledge in final products

The two following examples highlight how a domain Infrastructure and Architecture form an enterprise management Infrastructure. This global as-is state can be then redeployed through an enterprise management Architecture to ensure a better use of enterprise knowledge.

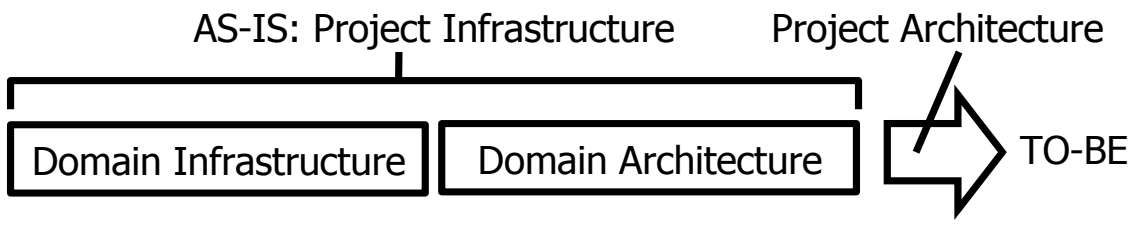

Figure 1: Two levels of Infrastructure and Architecture.

## 3 EXAMPLE ANALYSIS IN INSURANCE INDUSTRY DOMAIN

This example focuses on complex contract documentation found in insurance companies. It illustrates how the methodology introduced in this paper helped to analyse and improve the current Master Contract between the insurance company and its policyholders by first getting a good understanding of the current contract and its impact on the enterprise infrastructure. From this, improvements to the existing enterprise infrastructure could be made, and a new improved contract could be generated. This contract can then become a specification base for a CRM tool.

### 3.1 Context presentation

A given insurance company offers health cover, in certain categories, to its clients in the form of policies. A Master Contract governs all benefits and requirements between the insurance company and its policyholders. This contract is complex in nature mainly because of the use of complicated, inconsistent terminology, non-optimal grouping of related concepts, complicated product/business rule descriptions, ambiguous formulation, etc.

Many of these problems exist because different experts, from different domains within the company, are required to work on the same contract document. Matters are further complicated because experts don't always have a shared understanding, and no standard, agreed-upon terminology exists. Historically the contract also has evolved over a period of more than 60 years, adding to its Infrastructure intricateness.

In the event of a claim, the contract may in some cases be interpreted differently by the client and the insurer, leading to disputes between the two parties. Disputes damage the client-insurer relationship and may lead to extensive legal expenses for both parties. IT systems are used to calculate premiums, cover, claim payments, subject to certain rules and requirements. The complexity manifesting in the contract, also leads to complexity in the IT systems. In order to configure the IT systems, the contract must be interpreted. Ambiguous interpretations may lead to inconsistencies between the contractual terms and the IT systems, creating a legal risk to the company. In order to effectively, manage and reduce these complexities, a three-layered approach is followed.

1. Model and analyse contract paragraphs
2. Concepts are extracted from these contract paragraphs and relationships are established to other enterprise concepts

contained in an overall enterprise ontology.

3. Modelling of As-Is enterprise infrastructure, and developing of To-Be improvements.

The analysis of the paragraphs and the extraction of the paragraph concepts therefore correspond to understanding of the Domain Infrastructure. The Domain Architecture is then constructed by defining the relationships between these concepts and an enterprise ontology. This leads to a better understanding of the enterprise infrastructure (As-Is state), from which improvements can be made through an enterprise architecture (To-Be state). The process is summarised in Figure 2.

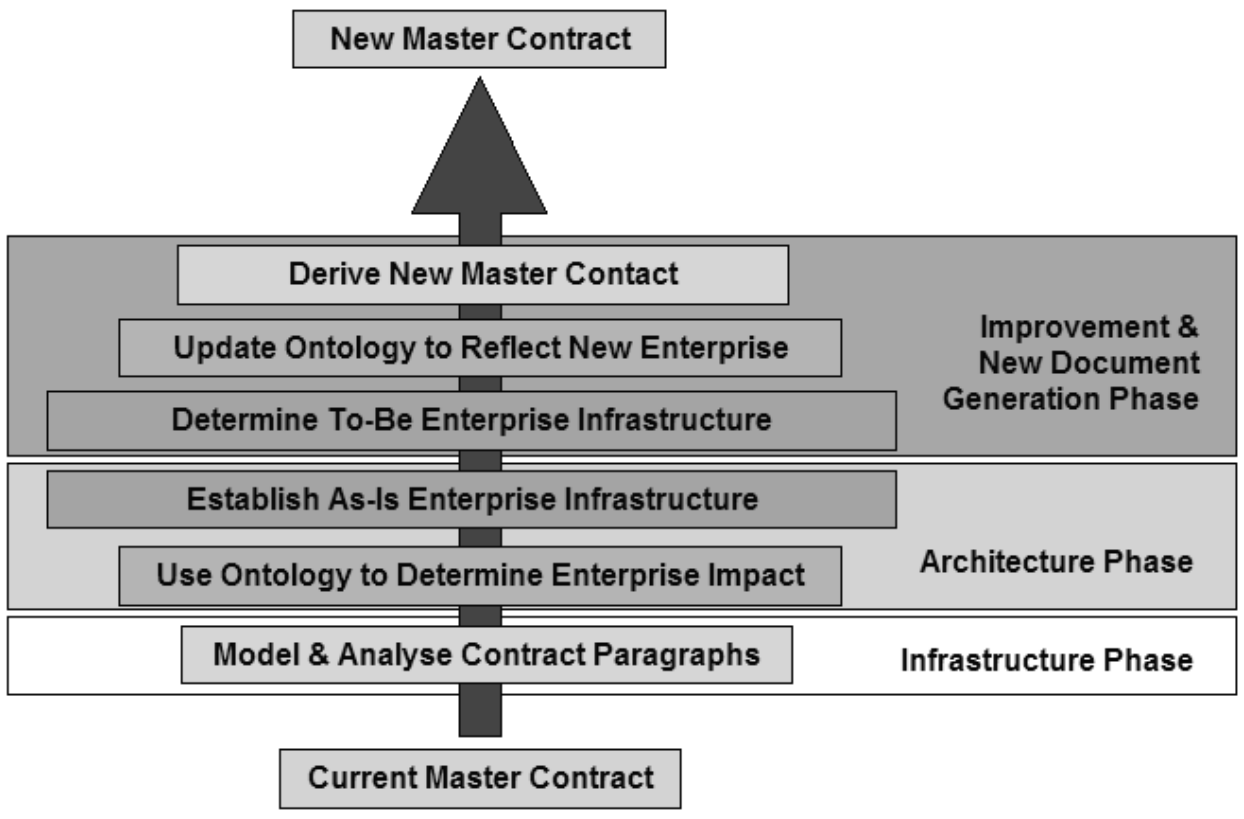

Figure 2 : Master Contract Improvement Process.

### 3.2 Contract infrastructure definition phase

Paragraphs are the basic building blocks of a document. In a legal document, paragraphs usually represent contract clauses that contain Role-players, Documents, Rules (Requirements / Restrictions), Events and Procedures. Because of non-coherent concepts used in paragraphs, complexities arise that lead to misinterpretation and misunderstanding.

As a first step to understand the existing contract infrastructure, paragraphs in the contract are analysed on a business process level by mapping flows within the paragraphs using a process-modelling tool called Moogo. This gives a visual representation of the interactions between the abovementioned elements in the paragraph and helps to overcome the interpretation limitations caused by natural language. Furthermore, interactions also exist between different paragraphs. Mapping the interactions in a mindmap format can highlight these interdependencies. This contributes to a better understanding of the influences that possible changes to contract paragraphs may have on other contract paragraphs. This understanding of the content and intent of paragraphs is an important first step in analysing complex documents for simplification.

### 3.3 Contract and enterprise architecture development phase

The next step is to extract specific concepts from these contract paragraphs and to establish relationships to other enterprise concepts contained in overall enterprise ontology. The impact of these paragraph concepts on enterprise concepts such as organisational structure, business processes, products, IT systems, the legal and legislative environment and allianced enterprises, etc., are therefore established and captured in an ontology. An ontology visualisation tool is used to portray various concepts and the relationships that hold among them. The resulting ontology helps to identify incoherencies in concept levels or typology definitions, as well as indicating what enterprise concepts are impacted by the Master Contract. In order to get a more detailed understanding of the impact on the enterprise architecture as a whole, the “as-is” state of the enterprise can be modelled using appropriate modelling tools as well as an enterprise modelling architecture (Modified PERA [16] in this case). Improvements in the business functions, information systems architecture, organisational architecture and hardware architecture can subsequently be developed using a comprehensive enterprise design life cycle architecture, thereby creating a required “to-be” state of the enterprise. Figure 3 presents an example of the PERA Master Plan architecture that serve as a roadmap for the improvement effort. An enterprise-wide innovation management software tool called EDEN™ supports this improvement effort. EDEN™ is a software environment, which provides high-level control over change-projects, supporting a multi-disciplinary team through a predefined roadmap structure [17]. It contains different enterprise reference architectures such as an enterprise life cycle, product life cycle, technology life cycle, and master plan for change projects that serves as a common reference framework and creating a unified understanding of the enterprise infrastructure. Good practice information is shared and experience and knowledge is captured within the EDEN™ environment. It ensures clarity, uniformity and coherency between different experts working on the same project. Figure 4 presents an example of the EDEN™ user interface.

### 3.4 Improved Contract Generation

In order to decrease the complexity and ambiguity of the Master Contract, it was necessary to firstly understand the contract, and secondly its impact on the enterprise as a whole. The root causes of the contract’s complexity and ambiguity lies not only in the contract itself, but also in the larger enterprise in the form of unclear/inconsistent business processes, complex product structures, ambiguous decision-making authorities and structures, It was therefore first required to improve the enterprise structure, and update the relevant ontology before an improved Master Contract could be generated. The enterprise-wide innovation management tool EDEN™ was used to support the collaborative modelling, improvement, and document generation processes.

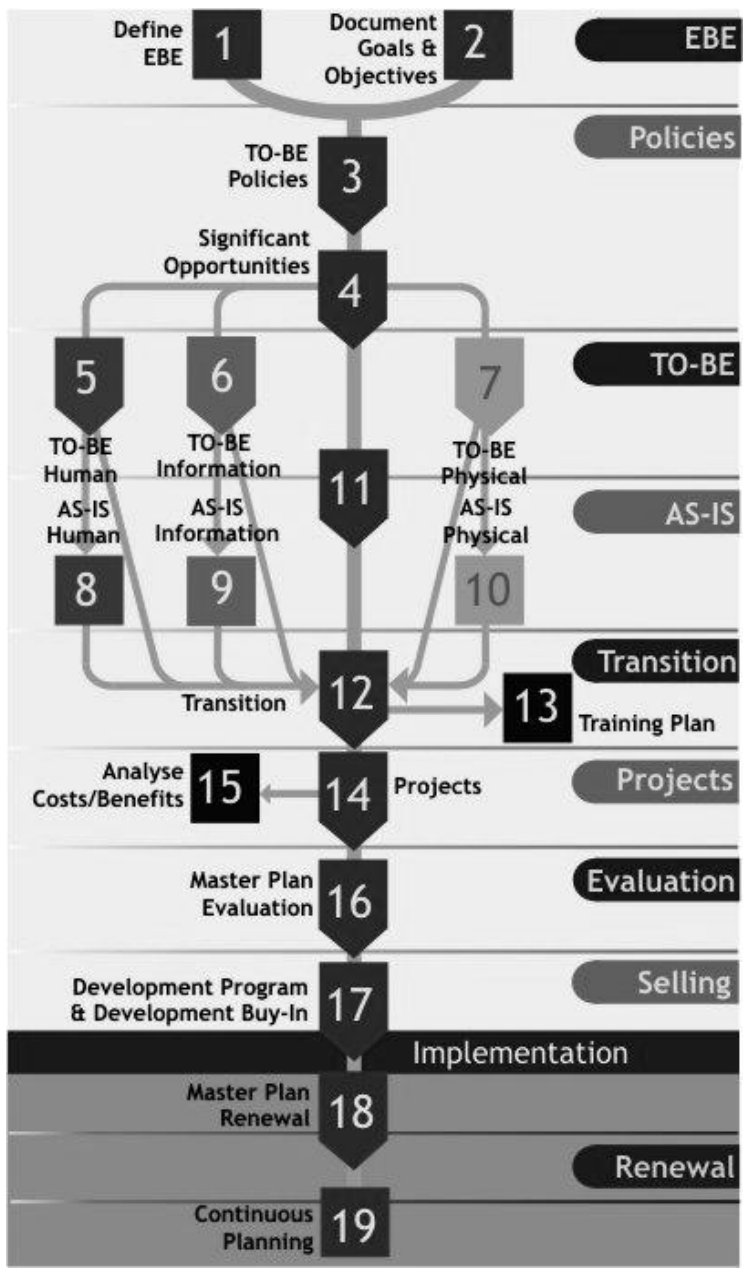

Figure 3: PERA Master Plan Roadmap.

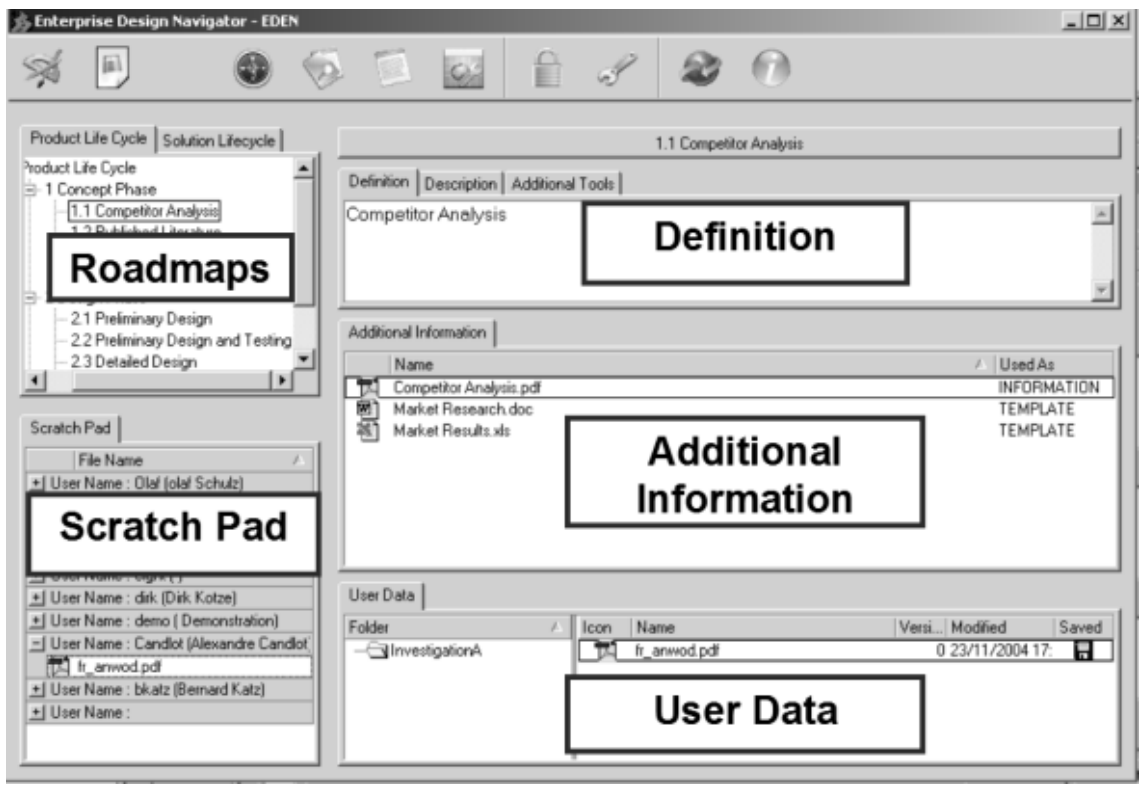

Figure 4: EDEN User Interface.

To summarise, ontology first help creating the as-is picture of the enterprise, giving an understanding of the enterprise, highlighting its incoherencies and guiding the to-be improvements formulation. In a second phase, the to-be changes to the enterprise have been implemented. Consequently, the first ontology had to be updated in order to incorporate the new evolutions. EDEN™ and roadmap methodology becomes a used common reference for developing a new, improved, simplified, coherent Master Contract that leads to more secured specifications of CRM tools.

## 4 EXAMPLE ANALYSIS IN AIRCRAFT MANUFACTURING DOMAIN

### 4.1 Definition of the couple Product / Process

In the world today, companies' computerization forces to assume that computer aided systems support design and manufacturing preparation phases [17]. Even if a global integration of the whole product and process life cycle is deployed, the harmonisation of the semantics associated to each expertise included in the life cycle remains difficult. When this integration issue isn't necessary for the enterprise, as in this example, several conceptual worlds keep on co-exist and forces multiple expert reinterpretations.

The following sections illustrate this case through the experience of an aircraft manufacturer. In a partnership with CAD/CAM development leader, five laboratories and a government organisation, the study of the possible integration of Manufacturing-Preparation Decision Process leaded to a Research Project supported by the French Ministry of Industry [19]. It aims at the specification and development of a knowledge-based engineering tool to help the definition of process plan for small-size high-specificity production batches.

Partner number and diversity have required a common reference to refine specifications from the expressed needs of the aircraft manufacturer to the final knowledge-based engineering tool. The process plan is the complex document considered here. It can be seen as the output of the KBE tool. This tool is specified either by the sum of diagrams (graphical representation) or by the sum of objects and their ties (informatics representation). Their evolution depends on two knowledge axes: the project maturity and the refinement of domain knowledge.

### 4.2 Project Infrastructure definition phase

Three main kinds of expertise were grouped on this project: aircraft manufacturing knowledge, scientific optimisation methods and CAD software programming. Each of them associates its own understanding to the following conceptual worlds: design specification, geometry and process capabilities.

The combination of these expertises (used for the project) and conceptual worlds (illustrating expertises contained in the life cycle) formed an implicit conceptual corpus. The first task of the Infrastructure Phase is to systematically identify them through documents, presentations and meetings. The justification of this work lies in the need of visibility and understanding. It creates an IS ontology [20] that represent the inter-subjectivity of stakeholders, preparing a coherent referring base to build the final KBE application.

A modelling tool, MEGA, helps to perform this systematic exploration. UML-like Activity diagrams and UML Class diagrams [21] [22] have been used and are presented on Figure 5 and Figure 6. The interest of these formalisms is that they propose a strong common syntax but let people free to rebuild their own semantic interpretation of meta-models. To keep advantage of the flexibility, this freedom nevertheless implies a clear common definition of this semantic before starting to model [23]. Each rework of all the models that can regularly occur when deep changes have been highlighted is the occasion to renegotiate the semantic interpretation. These reworks are significant leaps to maturity. During the project development, three principal phases appeared:

- Brainstorming phase: all partners propose the objects they feel they need for their part of the work. The separation of tasks is difficult and processes are often over detailed and over constrained. The number of concepts and processes rapidly increases.
- Homogeneity Phase: the knowledge on the existing software model increases and brainstormed concepts are progressively replaced by corresponding elements of the already-deployed models (Industrially or scientifically).
- Refinement phase: The tasks on processes are better distributed and are more homogeneous. The spine of the data model is progressively defined. Models are simplified by the identification of main concepts that are kept for rapid communication in the project team. The remaining items are semantically more relevant and coherent together.

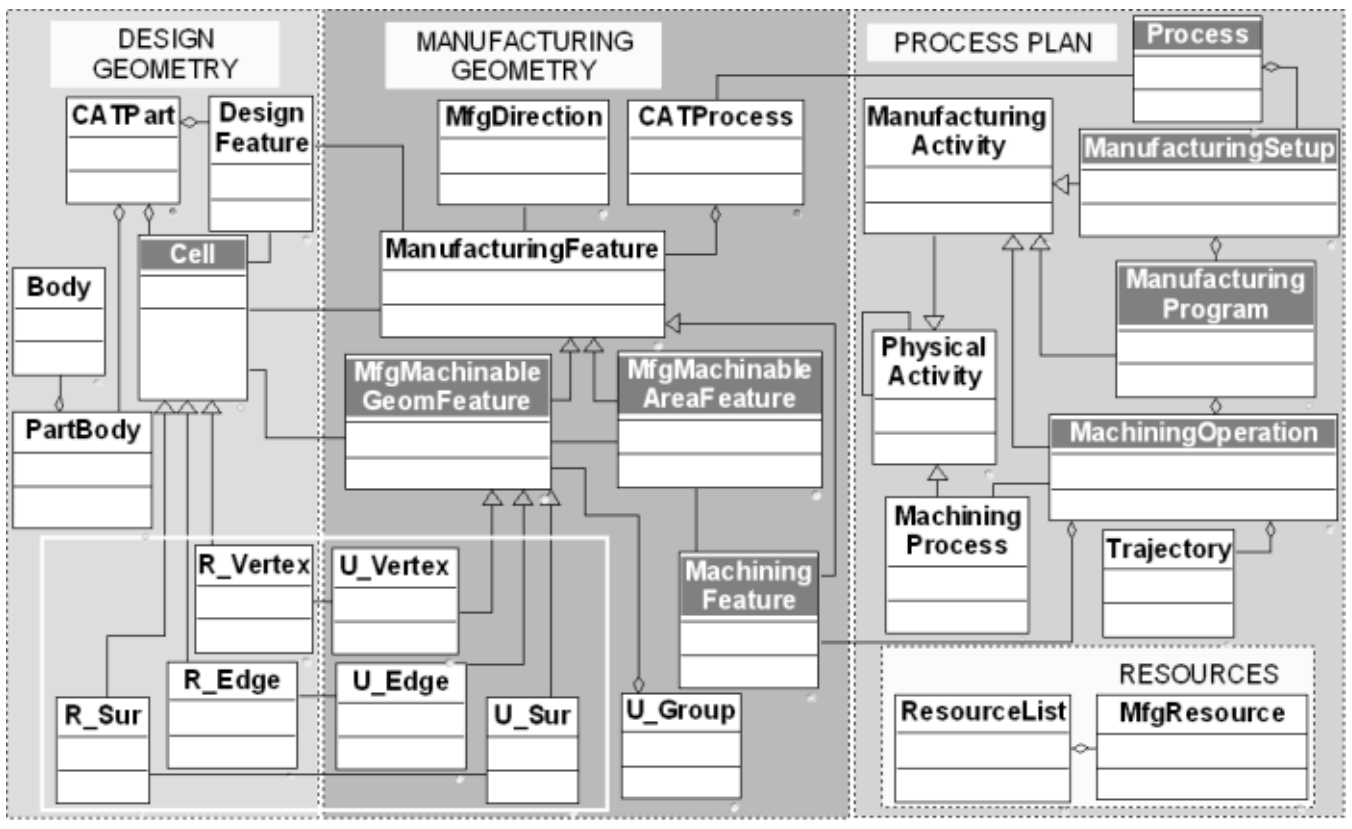

Figure 5: AS-IS Infrastructure Simplified Class Diagram.

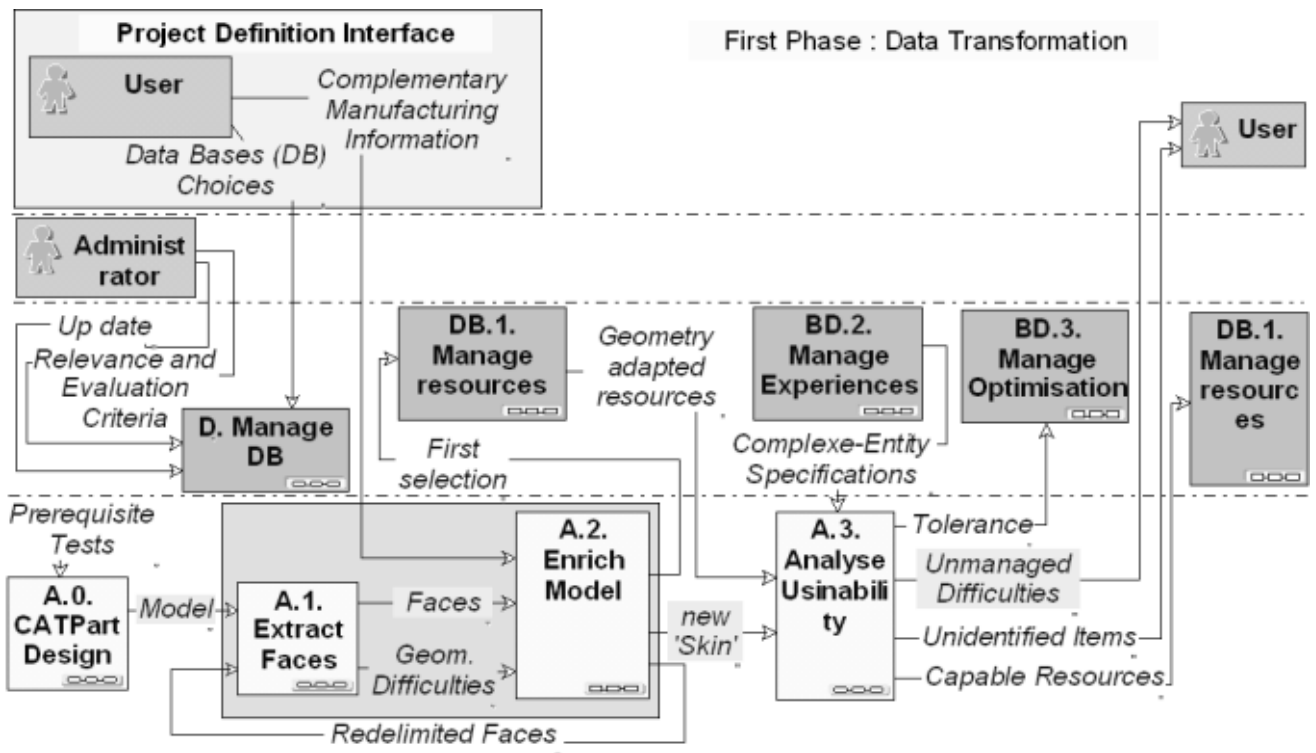

Figure 6: Part of the Domain Activity Diagram.

The last category is induced by the introduction of a temporal link between process model and data model. This link has been formalised by a sequence diagram. It creates a justification dynamic that can be illustrated by the Figure 7.

It must be noticed that this modelling method only helps to justify the need of elements but doesn't solve the definition of contained algorithms.

At this stage, the work on Project Infrastructure (concepts referring to process planning Infrastructure and Architecture) is combined to Architecture. The choices in scenarios become determining on the final platform efficiency.

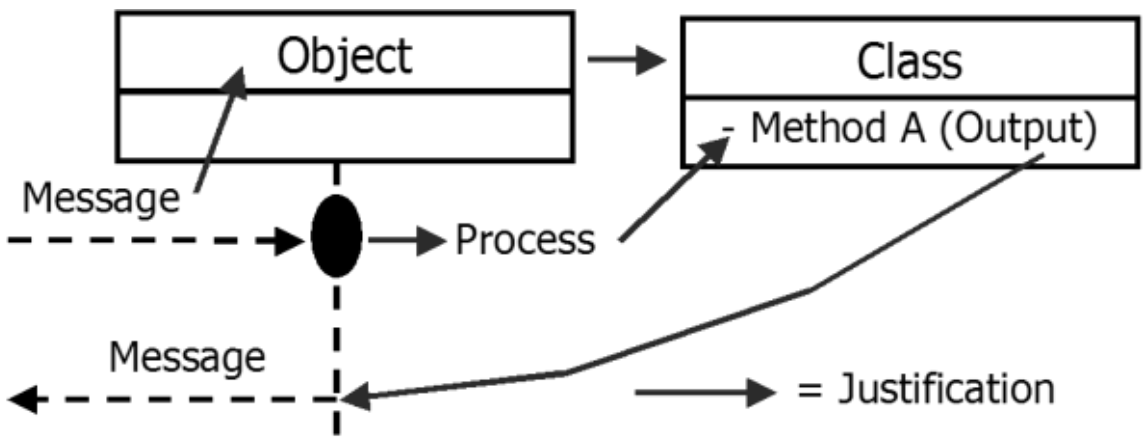

Figure 7: Concept Validation with Sequence Diagram.

### 4.3 Project Architecture constitution

As main items has been identified, the specification maturity process continues in detailing methods that make them interacting. In the presented industrial case, the specifications of the Architecture can be summed up to a need for a link between geometry and resources and a resolution of constraints problems

The central place of knowledge capitalisation through databases orientated the solution to an integration of Tools, possible manufacturing Sequences and classified Entities in one main database called OSE (French acronym for Tool, Sequence and Entities [24] – they are the high-value project-infrastructure concepts). For the constraint convergence of these semantically different items, a class system has been added to help the aggregation of all elements. The new Infrastructure organisation is presented on Figure 8.

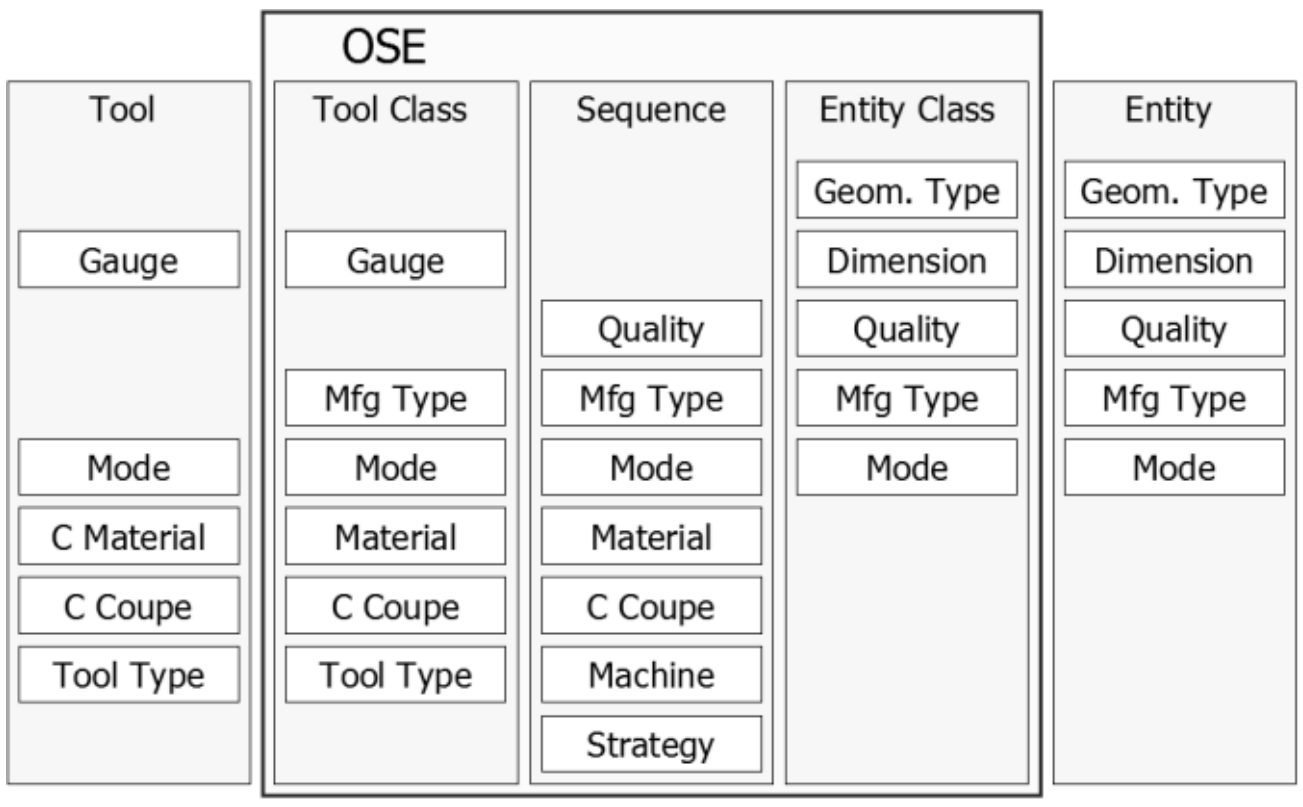

Figure 8: OSE Architecture Organisation.

The process flow using this solution appears in Figure 9. Two main steps can be highlighted. The first uses entity classes to map the studied mechanical part according to the formalised knowledge. The second uses tool classes to sort out relevant items and to open the structure to equipment evolutions. A last third step only consists in a classical cutting condition optimisation [17] and a validation of proposed solutions by the workshop.

### 4.4 Complex Document

Thus, for each new part considered, it becomes possible to first automatically identify the manufacturability of geometrical entities and then prepare the skeleton of the process plan by an organisation of sequences.

Another procedure helps to define the set-ups by calculations of accessibility directions. It hasn't been described here as on the contrary of the example, it involves homogenous mathematical data.

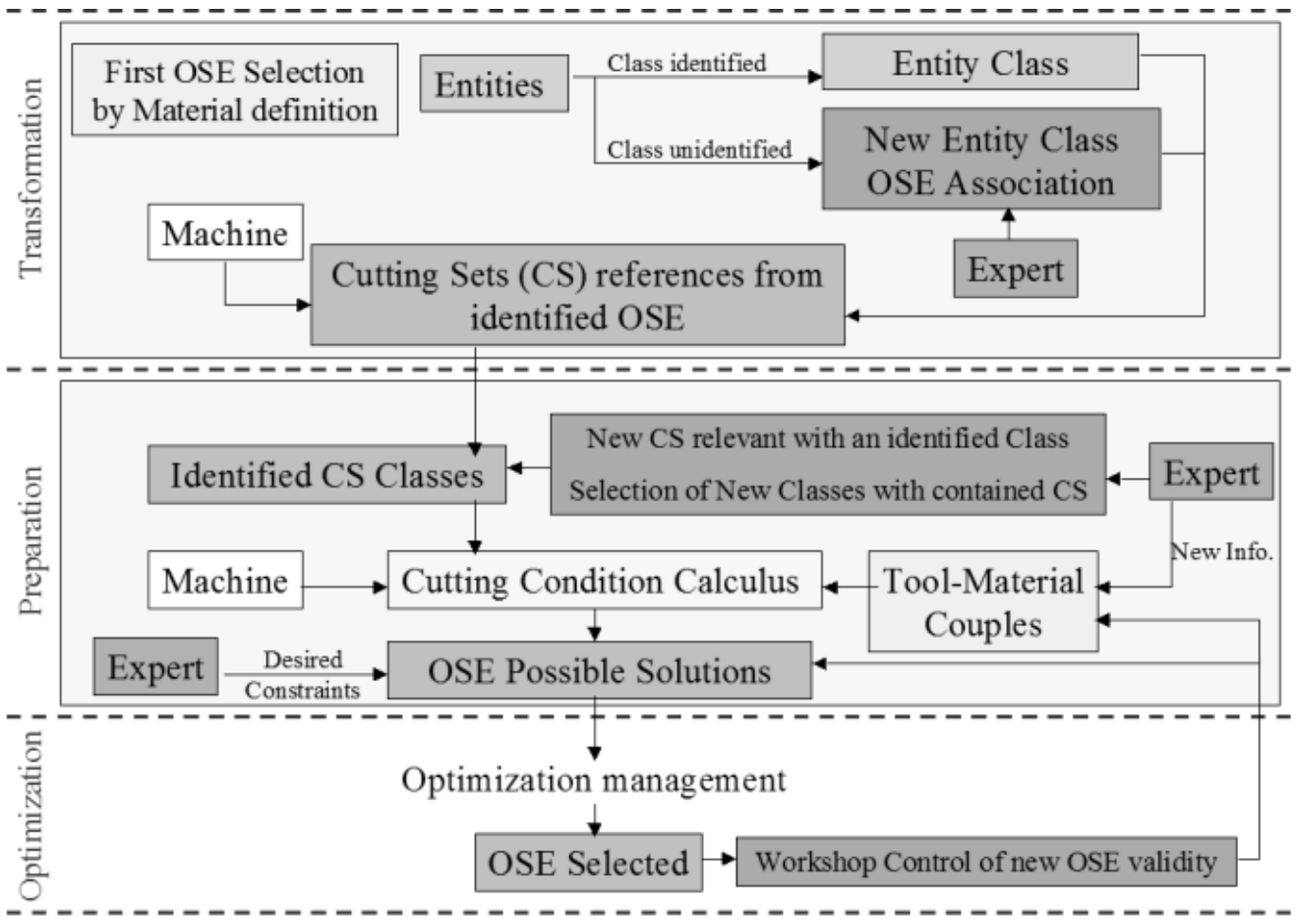

Figure 9: OSE Architecture Deployment.

With the help of these tools, the CAD document can rapidly be transformed in a CAM document, the expert concentrating his efforts only on difficult points. At the end of the pre-competitive phase of the development project, the expected gain is a ten factor. The resulting tool must still be completely deployed and validated by the industrialisation phase.

Other fields of application could expect benefits from modelling processes followed here. Cost optimisation in microelectronics components is prospected and a similar approach is deployed in this new context.

## 5 COMPARISON OF EXAMPLES

The conceptual similarities between these two very different application examples reminded on Table 1 encourage both teams to structure their work in a similar methodological approach.

| | Insurance Industry Example | Aircraft Manufacturing Example |
|---|---|---|
| **Domain Infrastructure & Architecture Specifications** | Master Contract | Expert Booklet |
| **Expected KBE Tool** | CRM Tool | Computer Aided Tool for Process Plan Definition |
| **Final Complexe Document** | Premiums, cover, claim payments | Process Plan |

Table 1: Example-Context Comparison

Naturally similar kinds of problematic are encountered, implying the search for common knowledge-based methods to solve them. It indicated that completely different tools (EDEN and Roadmap towards MEGA and UML diagrams) have deep similar concerns, i.e. to ensure coherency, consistency and a unified understanding in multi-faceted project teams.

The integration of both methodological aspects and technical solutions leads to a skeleton strategy reusable for further identical problems. The main experience outputs are summarised in Table 2.

In this process of specification refinement for a final complex document, strengths and weaknesses appear at the same stages in the two examples. The first drawback is that a relevant collection of terms is difficult to create. In both examples this task is manually performed. Informatics research works on algorithms to automatically analyse a document corpus in order to create a first cut of a structuring ontology.

The other difficulty lies in compiling statistical data in relevant indicators. The creation of a global methodology gives a better understanding of what should be monitored. It can be maturity phases or the respect of the methodological principles highlighted here. The Architecture phase goals are thus valuable hints for such indicators.

Next works should focus on how to monitor the global data evolution. Moreover the main advantage felt by the teams is the distinction between project and domain elements that is revealed in the need of clearly specifying the project syntaxes for a better construction of the domain semantics. The realisation of this problem is eased by the introduction of the Infrastructure and Architecture phases.

| Principles | Insurance Industry Example | Aircraft Manufacturing Example |
|---|---|---|
| Collect and Evaluate Information | Corpus Analysis / Meetings / Mindmaps | Corpus Analysis / Meetings |
| Sustainable Traceable Updating | Doc. Managemnt / Versioning System | MEGA Database Management |
| Accurate Overview | 3D Solution Space | MEGA Referential |
| Common Environment | EDEN Tools | MEGA |
| **Syntaxic Choices** | | |
| Data Representation | Metadata / Keywords | UML Class Standard |
| Process Representation | Moogo | UML Activity Standard |
| **Semantic Choices** | | |
| Domain Representation | Moogo Diagrams / Life Cycle Roadmap | UML Diagrams |
| Project Representation | Masterplan Roadmap (PERA) | UML Sequence Diagrams |
| Share Concepts between Users | Ontology Building | XMi / XML automatic generation |
| Data Analysis / Performance Indicators | Database Use Statistics | Referential Size Statistics |

Table 2: General to specific phase identification

## 6 CONCLUDING REMARKS

In a nutshell, this paper proposes a methodology to refine unorganised information complexity to semantically enriched relevant concepts. This reduction of the work structure complexity and heterogeneity helps to optimise the complex document generation by creating more coherent applications or work environments. The maturity of knowledge contained in these structures contributes to a better agility towards output expectations.

The introduced distinctions between Infrastructure and Architecture in one hand and Domain and Project in the other induce de facto an awareness of stakeholders on their position on a knowledge refinement scale. It avoids confusions in concepts considerations and allows the identification of global project risk.

The two specific contributions open ways to deploy these principles in different activity domains. The separation between methodology and methods comforts a step further the possible agility that today's enterprise project culture may require.